\documentclass[11pt]{article}
\pdfoutput=1

\usepackage{jheppub}

\usepackage{customprelude}

\newcommand{\be}{\begin{equation}}
\newcommand{\ee}{\end{equation}}
\newcommand{\bea}{\begin{eqnarray}}
\newcommand{\eea}{\end{eqnarray}}

\newcommand{\bln}{\begin{align}}
\newcommand{\eln}{\end{align}}
\newcommand{\bst}{\begin{split}}
\newcommand{\est}{\end{split}}
\newcommand{\bi}{\begin{itemize}}
\newcommand{\ei}{\end{itemize}}
\newcommand{\ben}{\begin{enumerate}}
\newcommand{\een}{\end{enumerate}}

\def\le{\left}
\def\ri{\right}

\def\al{{\alpha}}

\def\vev#1{\langle#1\rangle}

\def\th{{\theta}}

\newcommand{\p}{\partial}

\newcommand\ga{{\ensuremath{{\gamma}}}}

\newcommand\Lam{\Lambda}

\def\bR{\mathop{\bf R}}
\def\eeq{\end{equation}}

\newcommand\sM{{\ensuremath{{\mathcal M}}}}

\newcommand\sO{{\ensuremath{{\mathcal O}}}}

\newcommand\bpsi{{\bar \psi}}

\def\th{{\theta}}

\renewcommand{\todo}[1]{}

\title{A Goldstone theorem for continuous non-invertible symmetries}

\author{Iñaki García Etxebarria and}
\author{Nabil Iqbal}

\emailAdd{inaki.garcia-etxebarria@durham.ac.uk}
\emailAdd{nabil.iqbal@durham.ac.uk}

\affiliation{Department of Mathematical Sciences, Durham University, Durham, DH1 3LE, United Kingdom}

\abstract{We study systems with an
  Adler-Bell-Jackiw anomaly in terms of non-invertible symmetry. We
  present a new kind of non-invertible charge defect where a key role
  is played by a local current operator localized on the defect. The
  charge defects are now labeled by elements of a continuous
  $U(1)$. We use this construction to prove an analogue of Goldstone's
  theorem for such non-invertible symmetries. We comment on possible
  applications to string theory.}
\begin{document}

\maketitle

\section{Introduction}
Global symmetries and their associated conservation laws are one of the most fundamental tools that we have for the quantitative understanding of nature. 
One of the many uses of conventional global symmetries is to characterize phases of systems and their low-energy dynamics. To take one example, whenever a conventional continuous global symmetry is spontaneously broken, Goldstone's theorem guarantees that a gapless mode is present in the spectrum, and the low-energy dynamics is essentially completely characterized by the pattern of symmetry breaking. 

Recently, our understanding of global symmetries has undergone
something of a renaissance. The idea that the existence of conserved
quantities can be recast in terms of the topological surface operators
who count the charge has led to powerful generalizations of the very
concept of ``symmetry''. Two such generalizations which will concern
us in this paper are those of {\it higher-form symmetries} -- i.e. the
symmetries associated with the conservation of extended objects
\cite{Gaiotto:2014kfa} -- and {\it non-invertible symmetries},
i.e. symmetries for which the charges do not obey a simple group
composition law
\cite{Frohlich:2009gb,Carqueville:2012dk,Brunner:2013xna,Bhardwaj:2017xup,Gaiotto:2019xmp,Heidenreich:2021xpr,Choi:2021kmx,Kaidi:2021xfk,
  Roumpedakis:2022aik,
  Bhardwaj:2022yxj,Arias-Tamargo:2022nlf,Choi:2022zal,Choi:2022jqy,
  Cordova:2022ieu,Kaidi:2022uux,Antinucci:2022eat,Bashmakov:2022jtl,Damia:2022rxw,Damia:2022bcd,Bhardwaj:2022lsg,Lin:2022xod,Bartsch:2022mpm,Apruzzi:2022rei,GarciaEtxebarria:2022vzq,Lu:2022ver,Heckman:2022muc,Niro:2022ctq,Kaidi:2022cpf,Mekareeya:2022spm,Antinucci:2022vyk,Giaccari:2022xgs,Bashmakov:2022uek,Cordova:2022fhg}. See
e.g. \cite{McGreevy:2022oyu} for a recent pedagogical review of some
of these developments.

In particular, consider a system (such as conventional QED with a single massless Dirac fermion) where a current $j^{A}$ is nonconserved due to an Adler-Bell-Jackiw anomaly \cite{Adler:1969gk,Bell:1969ts}, i.e.
\be
d \star j^{A} = \frac{1}{4\pi^2} F \wedge F
\ee
where the operator $F \wedge F$ on the right-hand side is constructed from a {\it dynamical} $U(1)$ photon. It has recently been shown that though the naive symmetry is explicitly broken by the anomaly, this is not a ``generic'' kind of breaking; instead the system can be understood in terms of a novel kind of non-invertible symmetry \cite{Choi:2022jqy,Cordova:2022ieu}. This precise characterization opens new doors for a non-perturbative understanding of systems exhibiting such an anomaly. 

In this work, we present a variation of this construction that permits us to prove a Goldstone
theorem for such systems; i.e. we prove using Euclidean partition
function techniques that if an operator $\sO$ charged under the
non-invertible symmetry has a non-zero vacuum expectation value, there
must exist a gapless mode in the spectrum. As an application, we show
how many of the massless fields of string theory can be understood as
Goldstone modes of spontaneously broken non-invertible symmetries;
this provides an alternative viewpoint for their masslessness.

\vskip 0.5cm 
{\bf Note added:} In the last stages of preparation of this paper, \cite{Karasik:2022kkq} appeared, where a charge defect similar to our \eqref{chargeop} (involving an extra scalar field on the defect) was constructed for different motivations. 

\section{Goldstone's theorem for non-invertible symmetries}

\subsection{Goldstone's theorem in Euclidean formulation} \label{sec:goldstone1} 
To orient ourselves, we begin by reviewing a simple reformulation of the usual Goldstone theorem in the language of Euclidean path integrals. This argument was first given in \cite{Hofman:2018lfz} to prove a Goldstone theorem for higher-form symmetries. 

\begin{figure}[h!]
\begin{center}
\includegraphics[scale=0.6]{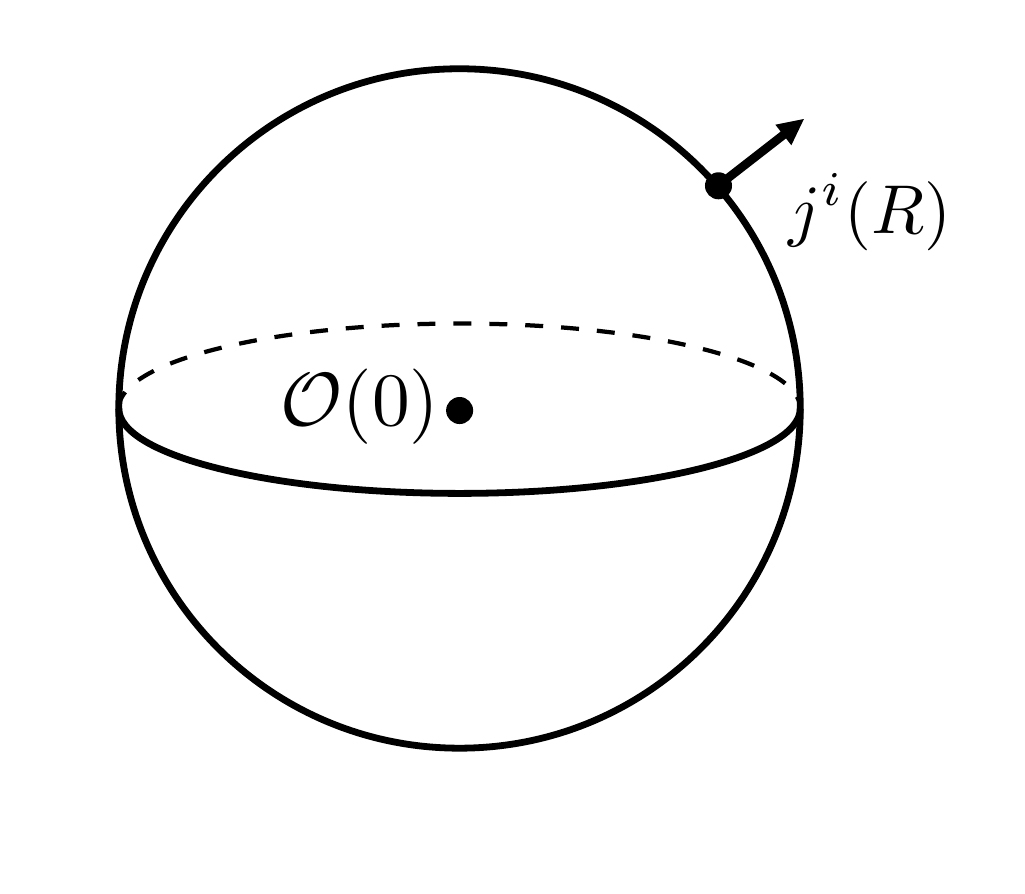}
\caption{\label{fig:sphere}
Charge operator defined on an $S^{3}$ of radius $R$ wrapping operator $\sO(0)$ at the origin.}
\end{center}
\end{figure}

Consider a Lorentz invariant quantum field theory with a $U(1)$ 0-form symmetry, with associated 1-form conserved current $j$. For simplicity we will restrict ourselves to the case of four spacetime dimensions. The Ward identity for the conserved current $j$ in the presence of a charged operator $\sO(x)$ with charge $q$ is
\be
d \star j(x) \sO(0) = i q \sO(0) \delta^{(4)}(x)
\ee
i.e. the current is not quite conserved in the presence of the charged operator. Let us now integrate both sides of this equation over a solid 4-ball of radius $R$ centered at the origin, as in Figure \ref{fig:sphere}. We find the equation
\be
\le(\int_{S^{3}(R)} \star j\ri)\sO(0) = i q \sO(0)
\ee
where on the left hand side we have used Stokes theorem; the integral is taken over the boundary of the 4-ball. Finally, take the expectation value of both sides:
\be
\bigg\langle\le(\int_{S^{3}(R)} \star j\ri)\sO(0)\bigg\rangle = i q \vev{\sO}
\ee

Now, if we are in a phase where the symmetry is spontaneously broken, then $\vev{\sO}$ is nonzero, and the integral on the left-hand side must be both nonzero and independent of the radius of the 3-sphere $R$. By spherical symmetry, we see that the correlation of the local operator $j$ on the 3-sphere and $\sO$ must therefore depend on $R$ as
\be
\langle j^{i}(x) \sO(0) \rangle \sim i q n^{i}  R^{-3} \label{simpcase} 
\ee
where $n^i$ is an outwardly pointing normal vector on the 3-sphere. The dependence on $R$ is fixed by the requirement that the integral over the 3-sphere result in an $R$-independent constant. Thus there is a power-law correlation in the theory.\footnote{This argument is of course exactly the same as the one used to obtain the inverse square electric field of a point charge using Gauss's law in elementary electrodynamics.} We have shown the existence of at least one gapless excitation: this is the Goldstone mode. 

We note that this is simply a reformulation of the usual Hamiltonian
arguments; if we appropriately deform the $S^{3}$ and cut open the
path integral then we obtain the commutators of $j$ and $\sO$ that are
used in the standard proofs. This Euclidean formulation will be useful
for the generalization that follows.

\subsection{Axial symmetry defect operators}

We now turn to the actual case of interest. In recent work
\cite{Choi:2022jqy,Cordova:2022ieu}, it has been shown that models
exhibiting the Adler-Bell-Jackiw anomaly actually are invariant under
a novel kind of non-invertible symmetry. We briefly review some
aspects of that discussion here.

Consider massless QED, defined by the action:
\be
S[\psi, \bpsi, A] = \int d^4x \le(\frac{1}{4 e^2} F^2 + i \bpsi \le(\slashed{\p} - i \slashed{A} \ri) \psi + \cdots \ri)
\ee \label{QED} 
The $U(1)$ gauge redundancy acts as
\be
\psi \to e^{i \Lam(x)} \psi \qquad A \to A + d \Lam \ . \label{U1gauge} 
\ee
In this theory the axial current $j_{A}^{\mu} = \bpsi \gamma^{5} \gamma^{\mu} \psi$ is not conserved. Instead, due to the ABJ anomaly it satisfies the following non-conservation equation \cite{Adler:1969gk,Bell:1969ts}: 
\be
d \star j^{A} = \frac{1}{4\pi^2} F \wedge F \label{anom} 
\ee
We stress that the right-hand-side of this expression is a dynamical operator, and not a fixed external source (as would be the case for a 't Hooft anomaly). There will thus be dynamical violation of axial charge conservation, and we cannot construct a conserved charge in the conventional manner. There is a temptation to instead consider the following current:
\be
\star j^{\mathrm{not-gauge-inv}}_{A} = \star j_{A} - \frac{1}{4\pi^2} A \wedge d A \ . 
\ee
This current is conserved, but as the notation suggests it is not gauge-invariant. Indeed it is not possible to construct a conserved gauge-invariant local current in this theory. However as the {\it integral} of the Chern-Simons term is gauge-invariant, one may try to construct a charge defect operator as follows:
\[
\hat{U}_{\al}(\sM_{3}) = \exp\le(i\frac{\alpha}{2}\int_{\sM_{3}}\le( \star j_{A} - \frac{1}{4\pi^2} A \wedge d A \ri)\ri) \label{Udef} 
\]
This is a topological operator under small deformations of $\sM_{3}$,
and thus is a candidate operator to define a conserved axial charge
defect, which acts on an operator $\sO$ with integer charge $q$ as
$\sO \to e^{i\alpha q/2} \sO$. 

However for general $\sM_{3}$ -- in particular those with nontrivial
1-cycles -- the expression \eqref{Udef} is not gauge-invariant under
{\it large} gauge transformations. Such an invariance would require
the coefficient of the Chern-Simons term to be quantized as
$\alpha \in 2\pi\mathbb{Z}$, thus resulting in a trivial action on all
operators. In \cite{Choi:2022jqy,Cordova:2022ieu} this deficiency was
remedied by adding extra degrees of freedom -- a TQFT -- living on
$\sM_{3}$. Their construction then permits the topological charge
defect operator to be defined on any $\sM_{3}$, provided that the
rotation angle $\alpha$ is in $\mathbb{Q}/\bZ$. Thus rather than having a continuous $U(1)$ symmetry as
in the conventional case the defect operators are now labeled by
$\mathbb{Q}/\mathbb{Z}$.

In this work however, we seek to generalize the Goldstone theorem described above, and it is important that we have some notion of a local conserved current. Thus we perform a slightly different construction. Consider the following charge defect operator, defined on $3$-manifolds $\Sigma$:
\be
U_{\al}(\sM_{3}) = \int [\mathcal{D}\theta] \exp\le(i\frac{\al}{2} \int_{\sM_{3}} \star \tilde{j}_{A} \ri) \equiv \int [\mathcal{D}\theta] \exp\le(i\frac{\al}{2} \int_{\sM_{3}} \le(\star j_A - \frac{1}{4\pi^2} (A - d\theta) \wedge dA \ri)\ri) \label{chargeop} 
\ee
Here we have introduced a new degree of freedom; this is a compact scalar $\theta$ that lives only on the 3-manifold $\sM_{3}$, and which transforms under the gauge redundancy \eqref{U1gauge} as $\th \to \th + \Lam$. In the above expression we perform a path integral over $\th$. Thus the integrand $\star \tilde{j}_{A}$ is now locally gauge invariant, and there is no rationality constraint on the parameter $\al$, so $e^{i\al} \in U(1)$, and we have constructed a continuous $U(1)$ symmetry. We will show below it is non-invertible. 

It should be noted that $\th$ really has almost no dynamics; we will restrict attention only to the case where $\sM_{3}$ is $S^{3}$, and in that case this construction can be viewed as a convenient way to extract the gauge-invariant information that is present in the more naive construction \eqref{Udef}. 

We now discuss some properties of the defect operator $U_{\al}(\sM_{3})$. 
\ben
\item It is topological under small variations of $\sM_{3}$. This can be seen heuristically by imagining some extension of $\th$ off of the defect, and then noting that $d\star\tilde{j}^{A} = 0$ for any such extension. In Appendix \ref{app:topinv} we provide a somewhat more detailed explanation of this fact. 
\item As usual for a 0-form defect operator, when the defect operator is collapsed onto an appropriately charged local operator it performs an axial rotation by an angle $\alpha$, i.e. if $\sO(x)$ carries axial charge $q$ then
\be
U_{\al}(S^{3}) \sO(x) = e^{\frac{i\alpha q}{2}} \sO(x) \label{charge1} 
\ee
if the $S^{3}$ in question is a small sphere wrapping $x$. 
\item We may immediately imagine interesting phenomena whenever $\th$ is allowed to have non-trivial winding, i.e. when (a) $\sM_{3}$ is a manifold with non-trivial 1-cycles, or (b) in the presence of 't Hooft lines. We discuss some such properties -- which do not affect our Goldstone proof --  in a subsequent section. 

\een

\subsection{Goldstone's theorem for non-invertible symmetry}
We may now use this object to prove a variant of Goldstone's theorem by generalizing slightly the arguments used in Section \ref{sec:goldstone1}. We begin by considering the setup of \eqref{charge1}, where $U_{\al}$ is defined on an $S^{3}$ of radius $R$. Now take a derivative with respect to $\al$ and set $\al$ to $0$: we then find
\be
\bigg\langle i \le(\int_{S^{3}(R)} \star \tilde{j}_{A}\ri) \sO(0) \bigg\rangle = iq \langle \sO \rangle
\ee
Note that the expectation value on the left-hand side now involves the path integral over $\th$ defined on the defect as well. $\tilde{j}_{A}$ is now defined only on the defect. 

Now imagine that we are in a phase where $\vev{\sO} \neq 0$, i.e. an operator charged under the non-invertible symmetry has a non-vanishing expectation value. Then, just as in \eqref{simpcase}, by spherical symmetry we see that the correlation function must depend on $R$ in a very specific way: 
\be
\langle \tilde{j}^i_{A}(R) \sO(0) \rangle \sim iq n^{i} R^{-3} \label{gaplessmode} 
\ee
with $n^i$ an outwards pointing normal vector on the $S^{3}$. We thus see that there is a non-trivial power-law correlation between an operator $\sO$ in the bulk and a current $\tilde{j}_{A}$ defined on the defect. Thus there must exist at least one massless mode in the bulk, which is the Goldstone mode. 

In this formulation the proof is essentially the same as in the
conventional invertible symmetry case, except that the current
$\tilde{j}_{A}$ is now defined only on the defect. Note that there is
no way for the new degree of freedom $\th$ living only on the defect
to create such a correlation between bulk and defect operators unless
the bulk is gapless.\footnote{It might be helpful to note that the
  expression above coincides with the \emph{connected} correlator, as
  the expectation value of the defect-localized current \emph{without}
  a local operator inserted is $\vev{\tilde{j}_{A}} = 0$, by the same
  argument as above with $q = 0$. Thus the connected correlator
  displays power-law behavior, requiring a gapless mode:
  \begin{equation}
    \langle
    \tilde{j}^{A}(R) \sO(0) \rangle - \vev{\tilde{j}^{A}(R)}\vev{\sO}
    \sim i q R^{-3}\, .
  \end{equation}}  This completes the proof.

The effective theory describing this low energy Goldstone mode (usually called an axion) is well understood. Denoting the axion by $\phi$, the first few terms in the derivative expansion are: 
\be
S[A,\phi] = \int_{\mathbb{R}^4} \le(\frac{1}{2\ga^2} d\phi \wedge \star  d\phi + \frac{1}{2e^2} F \wedge \star F + ig \phi F \wedge F\ri) \label{axion} 
\ee
It was shown in \cite{Choi:2022jqy} that this theory does indeed exhibit the non-invertible symmetry if $g = \frac{1}{4\pi^2}$. It is easy to verify that the $\phi$ field saturates the Ward identity \eqref{gaplessmode}, where the charged operator $\sO$ is realized as $\sO \sim e^{iq\phi}$.

An example of a microscopic theory which realizes this phase is if one
adds a complex scalar $\Phi(x)$ and a Yukawa coupling
$h \Phi(x) \bpsi \psi + h.c.$ to the Dirac action \eqref{QED}, and
then arranges a potential $V(\Phi^{\dagger}\Phi)$ to condense the
scalar $\Phi$; $\phi(x)$ may then be viewed as the phase of $\Phi(x)$
(or equivalently the phase of the fermion condensate $\bpsi\psi$).

Another example (discussed in \cite{Choi:2022jqy}) is QCD with massless quarks, where the (non-invertible) axial symmetry is spontaneously broken, and indeed the anomaly famously provides the dominant channel for pion decay to two photons. We now see that the $\pi_0$ can be viewed as a Goldstone boson of this breaking; the effective Lagrangian takes the form above with $\phi$ identified with the $\pi_0$ (and with other interactions involving the photon $A$).

We make two statements about generalizations of this Goldstone theorem:
\ben \item It may appear that the construction above requires access to a weakly coupled photon description of the theory. However the essential information of the ABJ anomaly can be phrased in a universal way\footnote{See e.g. \cite{Das:2022auy} where a similar characterization was used to understand finite-temperature physics from holography.} in terms of the combined symmetry structure of a conserved 2-form current $J$ and the non-conserved axial current $j^{A}$:  
\be
d\star j^A = \frac{1}{4\pi^2} J \wedge J \qquad d \star J = 0
\ee
We believe the argument can be extended to any theory with this structure. The second equation implies that we can locally write $J = \frac{1}{2\pi} \star dA$ in terms of an {\it effective} photon $A$. $A$ will not have a simple action, but we can nervertheless use it to construct the operator \eqref{chargeop} on a topologically trivial $S^{3}$ and run the argument above. 
\item Similar arguments can be made for a higher-form continuous non-invertible symmetry (e.g. as in \cite{Damia:2022rxw,Damia:2022bcd}) by changing the dimensionality of the defect manifold $\sM_{3}$ and of the charged operator $\sO$ (which will generically become an extended object), as was done in \cite{Lake:2018dqm,Hofman:2018lfz} for (invertible) higher-form symmetries. The statement then is that if the charged object exhibits a perimeter law (i.e. its expectation value depends only locally on geometric data characterising its worldvolume), then there exists a gapless mode in the spectrum.  
  \een
  
\subsection{Behaviour in the non-zero flux sector} 

For the proof above we were only required to construct the defect
operator $U_{\al}$ on an $S^{3}$ embedded inside $\mathbb{R}^{4}$. In
the interest of completeness, we would like to highlight some
properties of this defect on more general manifolds, particularly
those with non-trivial homology groups.

Let us consider formulating our bulk theory on $S^{1} \times S^{1} \times S^{2}$. We are interested in studying a sector of the path integral where there is $U(1)$ electromagnetic flux $2\pi n$ on the $S^{2}$, i.e.
\be
\int_{S^2} F = 2\pi n\, \label{magflux} 
\ee
where $n \in \mathbb{Z}$. 

Now let us place the defect operator $U_{\al}(\sM_3)$ on an $\sM_3 = S^1 \times S^2$, where it fills one of the two $S^{1}$'s, i.e. we evaluate

\[
   \langle U_{\al}(\sM_{3}) \rangle = \bigg\langle
   \exp\le(i\frac{\al}{2} \int_{\sM_{3}} \le(\star j^A -
   \frac{1}{4\pi^2} (A - d\theta) \wedge dA \ri)\ri)\, .
   \label{eq:U-correlator}
\]
We consider performing the path integral over $\theta$ first, keeping
$A$ fixed. Note that the dependence on $\th$ is only through its characteristic class (i.e. the winding), and any other $\th$ in the same class can be obtained by adding a topologically trivial $\th$ to a fixed representative of its characteristic class. We have
\[
  \int_{\cM_3} d\theta \wedge dA = 4\pi^2 w n
\]
with $w\in\bZ$ the winding of $\theta$ on the $S^1$ factor. Large
gauge transformations act as
\[
  G: (a, \lambda) \sim (a + n, w + n)
\]
where $a$ is the holonomy of $A$ on the $S^1$. In this calculation we treat $A$ as a background and we do
not impose these identifications when summing over $w$; in a more complete treatment they will be taken into account when performing the full
path integral over $A$ (where they reduce the effective range of the holonomy).  

The result of the path integral over $\theta$, including the winding
sectors, is therefore proportional to
\[
  \sum_{w\in\mathbb{Z}} \exp\le(i\frac{\al w n}{2} \ri)\, .
\]
Performing the sum over $w$, we find a delta function in $\al n$ that
only has support if
\[
\frac{\al n}{2} \in 2\pi \mathbb{Z} \ . 
\]
We thus see that for irrational $\frac{\alpha}{2\pi}$ our operator annihilates the path integral unless $n = 0$. This is a very non-invertible operator indeed. It would be interesting to further understand the properties of this defect and compare to the operator of \cite{Choi:2022jqy,Cordova:2022ieu}, which is  defined for rational $\frac{\alpha}{2\pi}$.

\section{Goldstone bosons and string theory}

One of our original motivations for seeking an extension of
Goldstone's theorem that applied to non-invertible symmetries was the
observation that in string theory there are fields which appear to be
exactly massless, even in non-supersymmetric configurations (as in
\cite{Alvarez-Gaume:1986ghj,Dixon:1986iz,Sagnotti:1995ga,Sagnotti:1996qj,Sugimoto:1999tx},
for instance). The masslessness of some of these fields is usually
explained from the fact that they are connections for higher-form
abelian gauge
symmetries. 

This argument is somewhat unsatisfactory on its own; a gauge
redundancy is after all a statement about a {\it description} of a
system. We should instead seek an explanation in terms of realizations
of (possibly approximate) {\it global} symmetries. Given our lack of
understanding of the non-perturbative aspects of string theory, a
systematic argument based on symmetry principles alone, valid for
arbitrary compactifications, seems to be of some value.\footnote{A
  different argument for the masslessness of the photon, based on
  swampland considerations, was given in \cite{Reece:2018zvv}.}

Consider, as an example, the operator measuring the D4 ``Page'' charge
\cite{Page:1983mke}\footnote{See \cite{Marolf:2000cb} for other
  notions of charge.}$^{,}$\footnote{The integrand
  $\tilde F_4 + A_1\wedge H_3$ can alternatively be written as $dA_3$,
  but the way we have written it makes it clearer that it is not gauge
  invariant under gauge transformations of $A_1$.}
\[
  \hat{U}_\alpha(\Sigma_4) = \exp\left(2\pi i \alpha \int_{\Sigma_4} \tilde F_4 + A_1\wedge H_3\right)\, ,
\]
where $\tilde F_4 \equiv dA_3 - A_1\wedge H_3$, $A_1$ and $A_3$ are RR
potentials, and $H\equiv dB_2$ is the NSNS field strength. Here
$\Sigma_4$ is a 4-surface linking a 5-manifold $X_5$ where we have
placed a D4 brane. We include no other background fields, and for
avoidance of tadpoles we take the directions transverse to the D4 to
be non-compact. 

Importantly, we will first treat the D4 brane as an infinitely
heavy object, so that we may consider it as a defect operator $D(X_5)$ in the effective supergravity field theory. We will eventually incorporate the fact that the tension is finite
below.

In this setting -- when the D4 branes are infinitely massive -- $\hat{U}_\alpha(\Sigma_4)$ is topological,
and would naively define a 5-form $U(1)$ symmetry. This is not so for
a number of reasons. First, for generic $\alpha$ the resulting
expression is not gauge invariant under large gauge transformations of
$A_1$ if $H^1(\Sigma_4;\mathbb{R})\neq 0$, and $H_3$ generic. We can
fix these issues by taking $\Sigma_4$ to be $S^4$. We also note that
the background we discuss does not source $H_3$, so we might be
tempted to simply claim that $U_\alpha(\Sigma_4)$ is gauge
invariant. But our goal is to argue that the presence of Goldstone
bosons is robust, so we will not set $H_3=0$. Instead, we introduce an
additional dynamical scalar on the charge defect, as above:
\[
  U_\alpha(\Sigma_4) = \exp\left(2\pi i \alpha \int_{\Sigma_4} \tilde F_4 + (A_1-d\theta)\wedge H_3\right)\, ,
\]
with $\theta$ transforming as $\theta\to\theta+\lambda$ under a
$A_1\to A_1+d\lambda$ gauge transformation. This defines a non-invertible symmetry analogous to \eqref{chargeop}.

The D4 brane is charged under $U_\alpha(\Sigma_4)$, so we may write:
\[
U_{\alpha}(\Sigma_4)D(X_5) = e^{i\alpha} D(X_5) \label{pagecharge} 
\]
Now the insertion of the defect operator $D(X_5)$, i.e. a D4 wrapping
$X_5$, in the string theory partition function will give a
contribution proportional to the volume of $X_5$, arising from the
Dirac-Born-Infeld coupling on the brane worldvolume. In terms of
standard language, it obeys a ``perimeter law'', so this nonzero
expectation value implies a form of spontaneous symmetry breaking. We
can, in particular, run a straightforward generalisation of the
argument given above (see \cite{Lake:2018dqm,Hofman:2018lfz} for
details) that implies the existence of a Goldstone mode. The minimal possibility saturating \eqref{pagecharge} is that we have a 5-form $C_5$ in the
spectrum, with a coupling
\[
  \exp\left(i\int_{X_5} C_5\right)
\]
to the D4 brane worldvolume. We identify this massless field as the
ordinary RR 5-form field.

Clearly, this argument generalises with minor modifications to all
branes in string theory, implying the exact masslessness of the fields
that they are charged under.

So far we have worked in a semi-classical approximation to string
theory, where the branes are infinitely massive objects. In actual
string theory, this is not the case: branes are dynamical objects, and
this means that there are no longer any exact global symmetries or
completely topological objects
\cite{Arkani-Hamed:2006emk,Harlow:2018tng,Banks:2010zn}. Nevertheless it seems reasonable that a large mass for these objects
should not affect questions in the extreme infrared, such as the
masslessness of putative Goldstone modes. An argument to this effect
for 1-form symmetries was given in \cite{Iqbal:2021rkn}. In our
context, the main physical effect is expected to be screening from
“virtual” D4-branes.  This makes the value of the measured Page charge
depend on the the distance $R$ between the surface where we measure
the charge and the brane, i.e. \eqref{pagecharge} becomes
\[
U_{\alpha}(\Sigma_4)D(X_5) = e^{i\alpha f(R)} D(X_5)
\]
for some nontrivial function of $R$. We are not aware of a systematic
treatment of this effect for the case at hand, but related cases are
analysed in detail in \cite{Cordova:2022rer}. Extrapolating the main
features of the cases discussed in that paper, we expect that the
operator we have described above will be topological --- i.e.
$f(R \to \infty)$ approaches a constant --- up to exponentially
suppressed effects, for scales larger than the inverse of the brane
tension.

This is good news: any small mass for the would be Goldstone would
dramatically affect our argument for large enough radii (the relevant
correlator would become exponentially suppressed), but it is precisely
in this regime that the screening effect itself becomes exponentially
suppressed. So we conclude that screening effects, while they do break
the symmetry, cannot give a mass to the Goldstone boson. This is
perhaps not much of a surprise: giving mass to a Goldstone boson for a
higher form symmetry, since it is vector-like, is a discontinuous
operation, so these Goldstone modes are much more robust than ordinary 0-form
Goldstone scalars (which have the same degrees of freedom whether they
are massive or not).

A summary of what we have just found is that the massless form fields
in string theory are massless because branes are heavy. This motivates
a final speculative comment: if we try to think what it would take to
gap the spectrum of supergravity, by the arguments above it will
necessarily involve making the branes tensionless. This resonates well
with the idea that gapped phases of gravity should be thought of as
phases where the metric vanishes. (We refer the reader to
\cite{Witten:1988ze,Witten:1988sy} and the final comments in
\cite{McGreevy:2016myw} for further references and discussion of such
ideas.) Indeed recent work on emergent higher-form symmetries
associated with the the gravitational sector alone
\cite{Hinterbichler:2022agn,Benedetti:2021lxj} may eventually allow us
to make such ideas concrete.

\acknowledgments

We thank Mohamed Anber, Ben Heidenreich, Arpit Das, Avner Karasik, Miguel Montero, Napat Poovuttikul, and
Tin Sulejmanpasic for discussions. Both authors
are supported by the STFC consolidated grant ST/T000708/1, and I.G.E
is additionally supported by the Simons Foundation via the Simons
Collaboration on Global Categorical Symmetries.

\begin{appendix}
\section{Topological invariance of defect operator} \label{app:topinv} 
Here we discuss the topological properties of the new coupling added to the action of the defect operator in \eqref{chargeop}, i.e.
\be
\exp\le(i\frac{\al}{8\pi^2}\int_{\sM_{3}} d\theta \wedge F\ri)
\ee
In particular, when placed in the path integral, is this invariant under deformations of $\sM_3$? 

Heuristically, if we extend $\th$ off of the defect, then we can consider deforming $\sM_{3}$ to a nearby surface $\sM_{3}'$; comparing the integral over the two regions we then find we then find 
\be
\int_{\sM_{3}} - \int_{\sM_{3}'} \le(d\theta \wedge F\ri)  = \int_{\sM_{4}}  d( d\theta \wedge F )  = 0 \label{topanswer} 
\ee
where $\p \sM_{4} = \sM_{3} \cup \sM_{3}'$. 

We can present a slightly more abstract formulation of the argument which shows the relationship between the above relation and the 1-form symmetry. Consider coupling an external 2-form source $b$ using the 1-form symmetry of QED:
\be
S[\psi, \bpsi, A; b] = \int d^4x \le(\frac{1}{4 e^2} F^2 + i \bpsi \le(\slashed{\p} - i \slashed{A} \ri) \psi + \frac{i}{2\pi} b \wedge F\ri)
\ee
(Such a coupling is possible for any conserved 2-form current $J$, where in this particular case we have $J \equiv \frac{1}{2\pi} F$). As $b$ couples to a conserved 2-form current, it is always true that
\be
Z[b] = Z[b + d\Lambda] \label{Zinv} 
\ee
for an arbitrary 1-form $\Lambda$. 

Now we turn on the following source for $b$, parametrized by a choice of 3-surface $\sM_{3}$: 

\be
b = b_{\sM_{3}} = \frac{\alpha}{4\pi} d \le( \Theta_{\sM_{3}}(x) d \theta(x)\ri) \label{bdef} 
\ee
Here $\th(x)$ is a scalar field defined on $\mathbb{R}^{4}$ which will eventually be restricted to the defect worldvolume $\sM_{3}$. $\Theta_{\sM_{3}}(x)$ is a scalar function with $\Theta_{\sM_{3}}(x) = 0$ if $x$ is ``on one side'' of $\sM_{3}$ and $\Theta_{\sM_{3}}(x) = 1$ if $x$ is on the ``other side'' of $\sM_{3}$. (E.g. if we were studying the simple case where $\sM_{3}$ is the flat plane $x_0 = 0$, we would have $\Theta_{\sM_{3}}(x) = \Theta(x_0)$ with $\Theta$ the usual Heaviside step function.) To make the manipulations well-defined, let us imagine that this interpolation from $0$ to $1$ takes place over a small length scale. 

Now as the choice for $b$ in \eqref{bdef} is exact, the invariance \eqref{Zinv} means that the partition function does not change under inserting this source. In particular it does not care about the choice of $\sM_{3}$, independently of the value of $\th$, i.e.
\be
Z[b_{\sM_{3}}] = Z[b_{\sM_{3}'}]
\ee
This invariance also does not care about details on how the Heaviside function above is regulated. However in the limit where the step function is made arbitrarily sharp, we find that $d\Theta_{\sM_{3}}$ becomes a 1-form localized on the defect, and we have: 
\be
\frac{i}{2\pi} \int_{\mathbb{R}^{4}} b_{\sM_{3}} \wedge F= \frac{i\alpha}{8\pi^2} \int_{\sM_{3}}  d\theta \wedge F
\ee
Thus we have constructed the coupling desired and confirmed that it is invariant under shifts of $\sM_{3}$. At the final stage we see that it depends only on $\th(x)$ on the defect. This argument is essentially the same as that leading to \eqref{topanswer}, but this presentation highlights the relationship between the topological character of the operator and the unbroken 1-form symmetry. 

\end{appendix} 
\bibliography{refs}
\bibliographystyle{JHEP}

\end{document}